\documentclass{mn2e}
\usepackage{epsfig}
\usepackage{times}
\begin{document}

\title[black holes] 
{The closest black holes}
 
\author[Fender, Maccarone \& Heywood]
       {R. P. Fender$^1$\thanks{email:r.fender@soton.ac.uk}, T. J. Maccarone$^1$, I. Heywood$^2$\\ $^1$ Physics and
         Astronomy, University of Southampton, Highfield, Southampton,
         SO17 1BJ, UK\\
       $^2$ Astrophysics, Department of Physics, University of Oxford, Keble Road, Oxford OX1 3RH, UK\\}
\maketitle
\begin{abstract}
Starting from the assumption that there is a large population ($\geq 10^8$) of isolated, stellar-mass black holes (IBH) distributed throughout our galaxy, we consider the detectable signatures of accretion from the interstellar medium (ISM) that may be associated with such a population. We simulate the nearby (radius 250 pc) part of this population, corresponding to the closest $\sim 35 000$ black holes, using current best estimates of the mass distribution of stellar mass black holes combined with two models for the velocity distribution of stellar-mass IBH which bracket likely possibilities. We distribute this population of objects appropriately within the different phases of the ISM and calculate the Bondi-Hoyle accretion rate, modified by a further dimensionless efficiency parameter $\lambda$. Assuming a simple prescription for radiatively inefficient accretion at low Eddington ratios, we calculate the X-ray luminosity of these objects, and similarly estimate the radio luminosity from relations found empirically for black holes accreting at low rates. 
The latter assumption depends crucially on whether or not the IBH accrete from the ISM in a manner which is axisymmetric enough to produce jets. Comparing the predicted X-ray fluxes with limits from hard X-ray surveys, we conclude that either the Bondi-Hoyle efficiency parameter $\lambda$ is rather small ($\leq 0.01$), the velocities of the IBH are rather high, or some combination of both.
The predicted radio flux densities correspond to a population of objects which, while below current survey limits, should be detectable with the Square Kilometre Array (SKA).
Converting the simulated space velocities into proper motions, we further demonstrate that such IBH could be identified as faint high proper motion radio sources in SKA surveys.
\end{abstract}

\section{Introduction}

Stellar-mass ($\la 15$M$_{\odot}$) black holes and neutron stars are the end points of
massive star evolution. The neutron star population we can investigate
both by their presence in binary systems (X-ray binaries or a small
number of pulsar--pulsar binaries) or as part of an isolated 
population, most evident as forms of radio pulsar (e.g Keane \& Kramer 2008 and references therein). 
The population of stellar mass black hole candidates (hereafter referred
to simply as black holes) is however less well understood, with fewer
than fifty well studied objects, all of which are in X-ray binary
systems (Lewin \& van der Klis 2006 and references therein).

Early estimates based on the star formation history of the Milky Way and the local mass density of stellar remnants put the total number of isolated, stellar-mass, black holes in our galaxy at
$\sim 10^8$ (e.g. Shapiro \& Teukolsky 1983; van den Heuvel 1992). This is broadly consistent with estimates of the rate of core collapse supernovae in our galaxy, and a ratio of total numbers of neutron stars to black holes in the range 3--10, depending on the slope of the IMF at high masses and the precise mass ranges for the formation of the two types of relativistic compact object. 

What fraction of these black holes are likely to be isolated and not in binary or multiple systems? 
Most O-stars are found in binaries, with indications of a preference
for massive companions rather than companions drawn from a standard
initial mass function (e.g. Pinsonneault \& Stanek 2006; Kobulnicky \&
Fryer 2007; Sana et al. 2012a,b).  Unbinding of a binary can take place
in several ways -- dynamical interactions; through mass loss during
the evolution of the more massive star; through the supernova
explosion of the initially more massive star; and through the
supernova explosion of the initially less massive star.  Additionally,
mergers of stars during binary evolution will produce single black
hole remnants from initially binary stars.

A rigorous calculation of the single fraction of stellar mass black
holes is beyond the scope of this paper, but we can show that $\sim
\frac{1}{2}$ or more of black holes will be single from fairly simple
considerations.  Sana et al. (2012a) find that about one quarter of
O-stars merge with their companions during their evolution.  They
additionally find that the mass ratio distribution for O star binaries
is consistent with being flat.  An initial mass of more than about 30
solar masses is required to produce a black hole, stars with
inital masses between 11 and 30 solar masses produce neutron stars
through the collapse of an iron core, while from about 8 to 11 solar
masses neutron stars may be produced through electron captures by
ONeMg cores (Fryer et al. 2012).  We can thus find that, assuming a
typical black hole progenitor mass of about 40 $M_\odot$, that about
half of black holes will have companions that will undergo iron core
collapse supernovae.  The neutron stars produced in these systems
should have natal kicks with typical velocities large enough to unbind 
any binaries with orbital periods long enough not to have undergone mergers.  
Between these two mechanisms, we can thus be confident, despite the 
crudity of our approach, that a large fraction of black holes, and probably 
a majority of black holes, will be single stars.

If we crudely model the Milky
Way as the combination of a disc component of thickness 0.5 kpc and
radius 10 kpc, combined with a spherical bulge of radius 2 kpc, the
total volume of the Galaxy is around 200 kpc$^3$. Assuming that $10^8$
isolated black holes (IBH) are
spread uniformly throughout this volume, there should be about $5
\times 10^5$ per kpc$^3$, with a mean separation of just over 10 pc.
This estimate is rather more conservative than that of Shapiro
\& Teukolsky (1983) by about 30\%, but significantly less so than Maccarone (2005), 
who considered a larger scale height for IBH. 
Dynamical friction (Chandrasekhar 1943) would cause this population to drift towards the Galactic
Centre, but this may not (yet) have strongly affected the population density
at the radius of Sun.
Note that nearest known strong black
hole candidate, in the binary A0620-00, is at a distance of about 1 kpc (Cantrell et
al. 2010 and references therein).

This large population of isolated black holes (IBH) should all accrete
at some level from the local interstellar medium (ISM), as should the related, larger, population of isolated neutron stars (INS). Several
authors have considered how to find this population of weakly
accreting objects via their optical (e.g. McDowell 1985) or X-ray
(e.g. Bennett et al. 2002; Agol \& Kamionkowski 2002, Perna et al. 2003) radiation, or via microlensing
(e.g. Agol et al. 2002b, Sartore \& Treves 2010).  However, there is another possibility. A
decade ago it was established that the relation between X-ray and
radio luminosities from black hole X-ray binaries (BHXRBs) was
non-linear, with a form $L_X \propto L_{radio}^b$ where $0.6 \leq b
\leq 0.7$ over a range of six orders of magnitude in $L_X$, albeit
with sampling heavily weighted to the higher luminosities (Corbel et
al. 2000; Gallo, Fender \& Pooley 2003; Gallo et al. 2006; Corbel et al. 2012). In recent
years a more radio quiet branch has been found below the main
correlation at higher luminosities (e.g. Coriat et al. 2011; Gallo,
Miller \& Fender 2012), but this does not affect the analysis
presented here unless this lower branch becomes dominant (and possibly
even diverges) at low luminosities. The non-linear relation between
radio and X-ray luminosities clearly implies that below some accretion
rate, sources should be easier to detect in the radio band than via
their optical or X-ray emission. This idea was discussed in depth by
Maccarone (2005), who demonstrated that future radio telescopes should
be able to find significant numbers of IBH, and that the radio band was in fact
a more effective route than searching for them with X-ray telescopes.
This may be especially true for cases of accretion in the cold cores of
molecular clouds which would suffer strong absorption at optical and
X-ray wavelengths.

In this paper we perform a simulation of a large population of nearby
IBH in order to investigate the detectability in radio and X-rays of
this population. For the first time we also calculate the expected
proper motions of the IBH population, and compare them to end-to-end
simulations of future radio telescopes, to see if they could be
detectable this way.

\begin{figure}
\epsfig{file=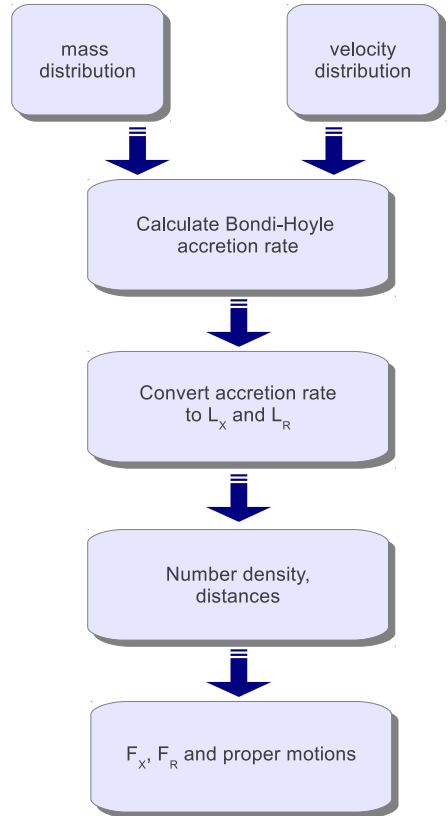,width=8cm}
\caption{A flowchart of the steps used in the simulation. The various
  steps of the simulation are described in detail in Section
  \ref{Sim}. The key points of uncertainty identified in the
  simulation were the velocity distribution of the black holes, and
  the efficiency of the Bondi-Hoyle accretion (parametrised by
  $\lambda$).}
\label{flowchart}
\end{figure}


\section{Simulation}
\label{Sim}

A numerical Monte Carlo simulation was performed of a sample of black holes
following the steps summarized in Fig \ref{flowchart} and described in
more detail in the subsections below. The simulation was of variable
size and tested repeatedly in order to refine our understanding of the
key unknowns. 

\subsection{Mass distribution}

The mass distribution of black holes in X-ray binary systems (XRBs) is
not well understood, but has been known for some time to have an
average value of around 7 M$_{\odot}$ (Bailyn et al. 1998). \"Ozel et
al. (2010) show that the distribution of 16 measured BH masses in XRBs
can be represented by a Gaussian distribution with mean of 7.8
M$_{\odot}$ and a standard deviation of 1.2 M$_{\odot}$. This
conclusion is consistent with a parallel study by Farr et
al. (2011). We use the \"Ozel et al. distribution in our simulation.

\subsection{Velocity distribution}

The distribution of 3D space velocities of IBH, $v_{\rm 3D}$, is
essentially unknown (unsurprisingly since no IBH has yet been
identified).  However, the velocity distribution of nearby stars is
relatively well measured, and is a function of spectral type, with
one dimensional velocity distribution
$\sigma_{\rm 1D} \sim 15$ km s$^{-1}$ for nearby early type ($M \geq 2 M_{\odot}$) stars, and $\sigma_{\rm 1D} \sim 40$ km s$^{-1}$ for nearby late type ($M \leq 1 M_{\odot}$) stars (e.g. Dehnen \& Binney
1998 and references therein). We consider the scenario in which the IBH have the velocity
dispersion of the nearby early-type stars as a lower limit to their
peculiar velocities.  This is referred to in the rest of the paper as
the `no kick' scenario.

We may also consider that the IBH receive a kick during the supernova
explosions associated with their formation. One way to do this is to
take the distribution measured for NS and extrapolate this to BH under
the assumption that the impulses given in the supernova explosion are
comparable. For NS, we follow Hobbs et al. (2005), in which the 3D
space velocity distribution for NS is given by a Maxwellian
distribution with scale parameter 265 km s$^{-1}$. In converting this
for the BH, taking a ratio of mean BH to NS masses of $7.8/1.4 = 5.6$,
we simply assume conservation of momentum and divide the NS velocity
distribution by 5.6 to get the BH velocity distribution.  This is
referred to in the rest of the paper as the `large kick' scenario.
For further discussions on observational constraints on black hole
kicks, see e.g. Mirabel \& Rodrigues (2003), Jonker \& Nelemans
(2004), Gualandris et al. (2005); Miller-Jones et al. (2009). 

\begin{figure}
\epsfig{file=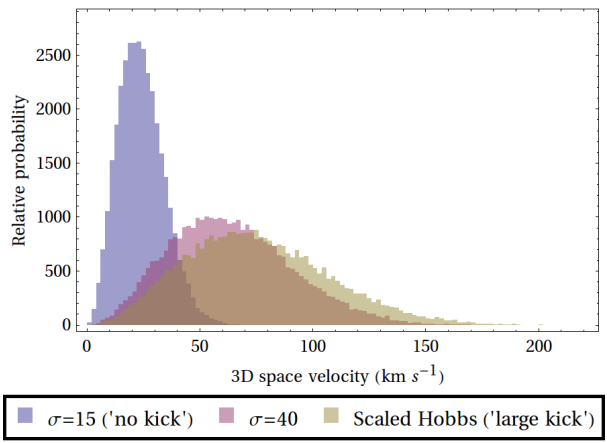,width=8cm}
\caption{Three velocity distributions considered for the population of isolated black holes (for a single realisation of the simulation). At the lower end, we consider a Normal (Gaussian) distribution of velocities with velocity dispersion 15 km s$^{-1}$, at the upper end we consider a velocity dispersion of 40 km s$^{-1}$ as well as the Hobbs et al. (2005) velocity distribution for radio pulsars divided by the mean IBH:NS mass ratio of 5.6. Other distributions are of course possible but probably fall in the range covered by these extremes.}
\label{velplot}
\end{figure}

Note that there are other, differing, descriptions of the NS velocity
distribution, such as that put forward by Arzoumanian, Chernoff \&
Cordes (2002). In their distribution there is a larger number of low
velocity neutron stars compared to the Hobbs et al. (2005)
distribution. In terms of considering a wide range of velocity
distributions for the IBH population being simulated, this is in the
direction of the `no kick' scenario, and so we continue to consider
the Hobbs et al. distribution as the extreme for the `large kick'
scenario. Equally, considering the higher velocity dispersion of late
type stars shifts the results in the direction of the `large
kick' scenario, and so these two scenarios, `no kick' and `large kick', should bracket all the
likely possibilities (although of course we simply cannot exclude the possibility that
there are some very high velocity IBH). The velocity distributions for these scenarios are
plotted in Fig \ref{velplot}.

\subsection{Properties of the interstellar medium}

In this analysis we use the mean properties of the ISM within our
galaxy, derived from those tabulated in Longair (1997; acknowledged as
originating with Dr John Richer). We consider four components of the
ISM, with their corresponding volume fraction, mass fraction, density,
temperature and sound speed. The sound speed $c_s$ is calculated under
the assumption that, for gases in pressure balance, $c_s \propto T^{1/2}$, where $T$ is the gas
temperature. The mean properties of these
four components are given in table \ref{ism}.  For each component we actually simulate a range
of values corresponding to a Normal distribution centred on the mean and with a dispersion of 20\%. 
However the effect of these broader ISM  distributions is in fact almost completely dominated by the
distribution of black hole velocities when calculating the Bondi accretion rate.

\begin{table*}
\begin{tabular}{cccccc}
\hline
Phase & Volume & Mass & N & T & c$_s$ \\
& Fraction & Fraction &  (cm$^{-3}$) & (K) &  ($\times 10^5$ cm s$^{-1}$) \\
Molecular Clouds & 0.005 & 0.4 & 1000 & 10--30 & 0.6 \\
Diffuse Clouds & 0.05 & 0.4 & 100 & 80 & 0.9 \\
Intercloud Medium & 0.4 & 0.2 & 1 & 8000 & 9 \\
Coronal Gas & 0.5 & 0.001 & 0.001 & $10^6$ & 100 \\
\hline
\end{tabular}
\caption{The four main phases of the ISM, simplified. The volume
  filling factor, mean number density, and mean sound speed are used directly in
  the simulations, in which the properties of each each phase of the ISM are calculated
  assuming a normal (Gaussian) distribution based on the means, with a 20\% dispersion.
  The tabulated values are simplified and derived from Longair (1997), who in turn
  attributes them to Richer.}
\label{ism}
\end{table*}

However, it is also widely accepted that the local ISM is in general relatively hot and tenuous on scales of
$\sim 70$ pc in all directions (Frisch, Redfield \& Slavin 2011 and
references therein), the so-called Local Hot Bubble (LHB). The LHB is not uniformly hot and tenuous, however,
and contains both warm and cold clouds (Frisch et al. 2011; Peek et al. 2011). Objects within this
bubble are of course very important for the proper motions part of this
investigation, and so we need to consider this aspect carefully.  

In fact the filling factor of cooler gas within the LHB is not well known, but on
scales between 15--70 pc is likely to be $\la 1$\% (Frisch, Redfield \& Slavin, {\em private communications}). However, very close to The Sun, within 15 pc, there is a local overconcentration of warm clouds, with a volume filling factor of $\sim 10$\%. Based on the number density estimated earlier, this volume of radius 15 pc should contain 5-10 IBH, which means that at most one should currently be accreting from these local clouds. 

In order to account for the LHB and the local warm clouds, it is simplest to assume that within a radius of 70 pc all the gas is in the coronal phase, but we note that there is a reasonable chance that one IBH may be accreting from one of the local warm clouds.

\subsection{Calculation of accretion rates}

With a distribution of BH masses and space velocities, combined with
densities and sound speeds of the four major phases of the ISM, we are
able to calculate the Bondi-Hoyle-Lyttleton accretion rates for the set of simulated
objects, using the following equation:

\[
\dot{m}_{\rm Bondi} = 4 \pi \lambda (G M_{BH})^2 \rho (v_{BH}^2 + c_s^2)^{-3/2}
\]

\noindent
which originates with Hoyle \& Lyttleton (1939) and Bondi \& Hoyle (1944), 
where $G$ is the
Gravitational constant, M$_{BH}$ is the BH mass, $\rho$ is the
density of the gas, $v_{BH}$ is the space velocity of the BH and $c_s$
is the sound speed of the accreting gas. The parameterisation of the
accretion efficiency relative to the Bondi-Hoyle rate, $\lambda$, is
rather hard to estimate. Agol \& Kamionkowski (2002a) assumed $\lambda
= 1$, however this is likely to be far too optimistic. As noted in Perna et al. (2003) and references therein, there would be a large population of bright X-ray sources corresponding to nearby isolated neutron stars if this were the case, 
but these are not observed. Perna et al. considered that $0.0001 \leq \lambda \leq 0.01$. In the related area of AGN accretion, Pellegrini (2005) concluded that effectively $\lambda \sim 0.01$.
In section \ref{results} below,
we use a combination of our simulation and current hard X-ray survey
limits to place further constraints on this.

\subsection{Radiative luminosity (X-ray emission)}

For radiatively efficient accretion at relatively high rates onto a BH, the radiative
efficiency $\eta \sim 0.1$, where luminosity is related to accretion rate as $L = \eta \dot{m} c^2$. This should be similar to that for NS. However,
it has been argued that at low accretion rates much of the available
accretion power is advected across the black hole event horizon in
hot, low density flows (Ichimaru 1977; Narayan \& Yi 1995). 

One commonly used formulation is to assume that the radiative
efficiency of black hole accretion decreases linearly with accretion
rate below some threshold at around 1\% of the Eddington accretion
rate.  This in turn leads to a quadratic dependence of accretion
luminosity on accretion rate for low luminosity sources, $L \propto
\dot{m}^2$. There is some observational evidence that this is
approximately the right scaling (Koerding, Fender \& Migliari 2006),
although it is far from conclusive.

In calculating the radiative efficiency of the BH in our simulation,
we therefore use the following formulation:

\[
\eta_{BH} = 0.1 (\dot{m} / \dot{m}_{crit})
\]

\noindent
where $\dot{m}_{crit}$ is the accretion rate corresponding to 1\% of
the Eddington luminosity for a 7 M$_{\odot}$ black hole accreting with
$\eta = 0.1$ ($1.1 \times 10^{17}$ g s$^{-1}$).

\subsection{Jets (radio emission)}

Fender (2001) established that all hard state accreting black holes
(which includes, but is not exclusive to, all BH observed at less than 
about 1\% of their Eddington luminosity) produce GHz radio
emission with a more or less flat spectrum (possibly extending
as low as 300 MHz and as high as the near-infrared band), indicative of a compact
partially self-absorbed jet like those originally postulated to
explain the radio cores of AGN (Blandford \& K\"onigl 1979). 

As discussed in the introduction, observed correlations between accretion
flows and GHz-frequency radio emission suggest a relatively simple scaling over many
orders of magnitude in accretion rate.

In many models of jet formation, it is assumed that -- unlike the
radiative output (see above) -- the jet kinetic power is linearly
proportional to the accretion rate (Meier 2003 and references
therein). K\"ording et al.  (2006) established an empirical relation
between this jet power and the radio luminosity of a source, which can
be simplified to:

\[
L_R \sim 10^{5} (\dot{m})^{17/12} \phantom{00} \rm {erg s^{-1}}
\]

\noindent
ignoring two terms of order unity (which we cannot in any case estimate more
accurately in this analysis). This and other relations estimated in
Koerding et al. (2006) are broadly consistent with other approaches to
the same problem based on studies of AGN (La Franca, Melini \& Fiore
2010 and references therein). We apply this formula to the calculated
accretion rates to derive the radio luminosities of the accreting
isolated black holes.

\subsection{Distances and fluxes}

Assuming that we are situated close
to the midplane of the Galactic disc, a simulation of a sphere centred
on us and extending to the `edges' of the disc (i.e. radius 250 pc),
should contain around $3.5 \times 10^3$ black holes. 
In order to calculate the observable fluxes from the simulation, we
simply assume that these IBH are randomly distributed within the sphere
of radius 250 pc centred on the Sun, and calculate the observed flux
densities. 

For the radio emission, we assume a relation between the radio luminosity
and the GHz spectral luminosity ($L_{\nu}$) based on a flat 
(spectral index $\alpha = 0$, where $S_{\nu} \propto \nu^{\alpha}$) radio spectrum:

\[
L_{\nu} = \frac{L_R}{\nu}
\]

For the X-ray emission, we need to convert between the total radiative luminosity, $L$, and 
the hard X-ray luminosity, $L_X$, which we use below for comparison with existing hard X-ray surveys.
Assuming that the total radiative luminosity is represented by a power law of photon index +1.6
(spectral index $\alpha = -0.6$), then the bolometric correction is approximately 0.3, where

\[
L_X = 0.3 L
\]

This is based upon a typical hard state black hole spectrum, which should be appropriate at low
accretion rates. The X-ray flux and radio flux density are simply calculated by assuming isotropic emission and dividing $L_X$ and $L_{\nu}$ respectively by $4 \pi d^2$.

\begin{figure*}
\epsfig{file=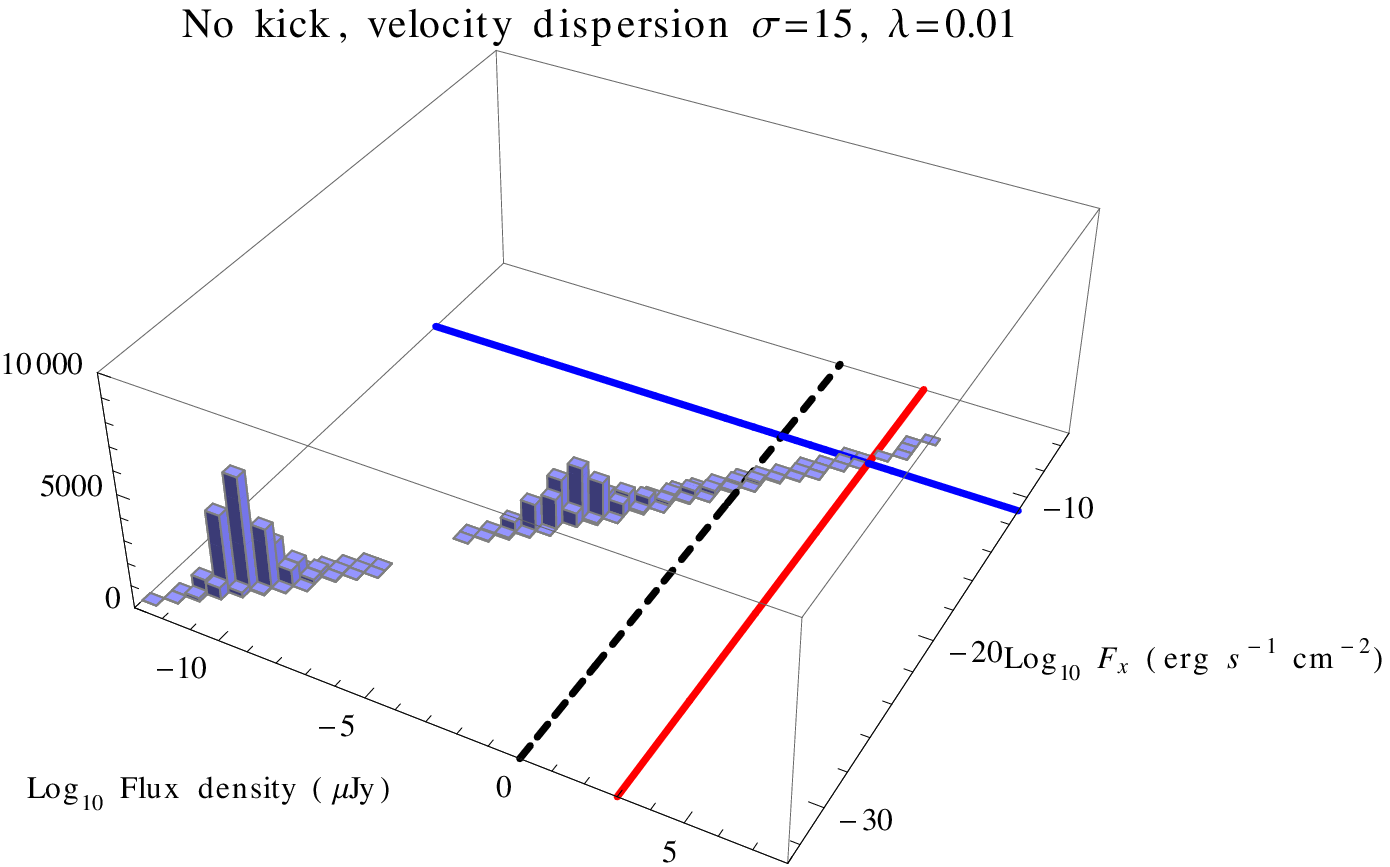,width=8cm}\quad\epsfig{file=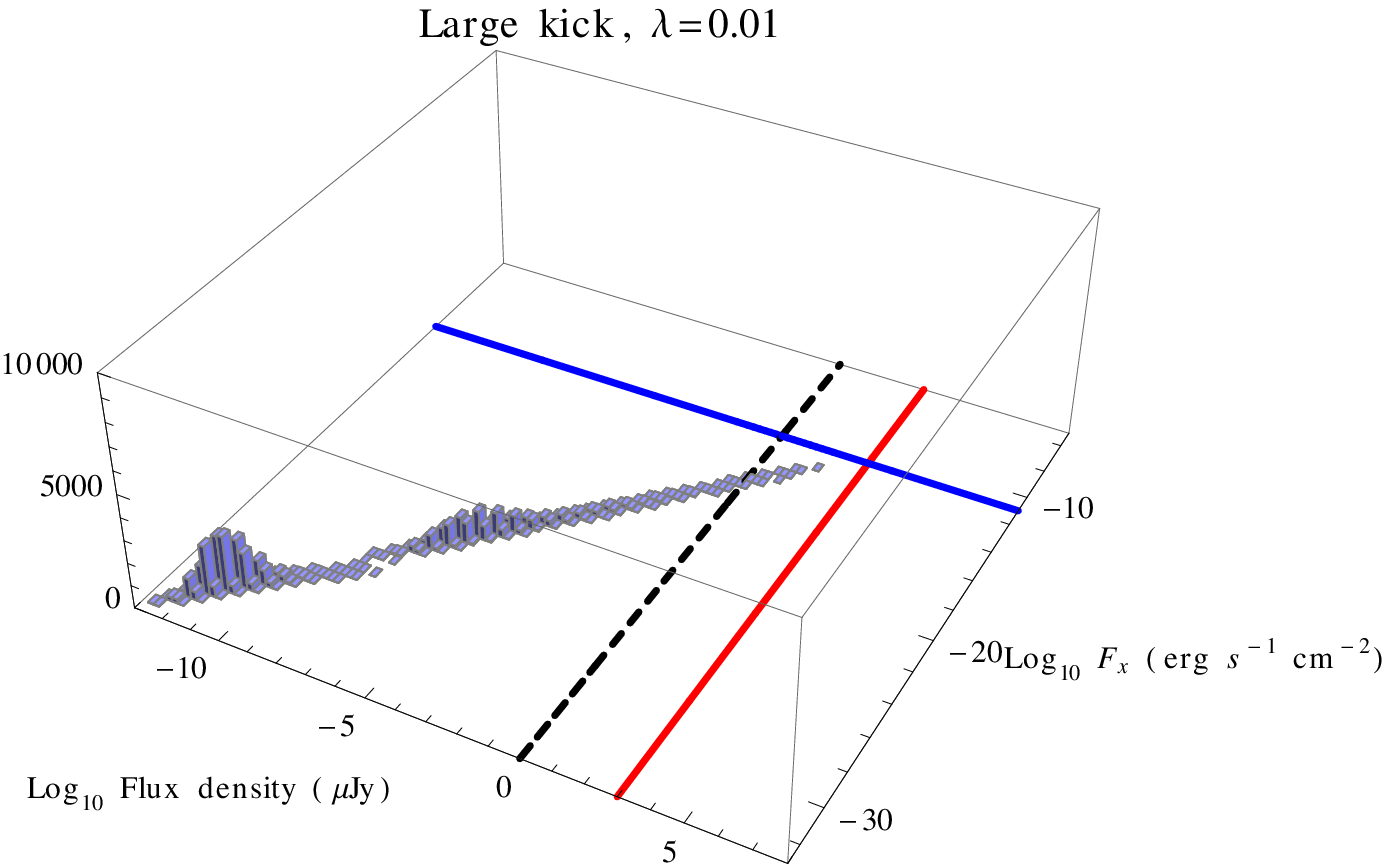,width=8cm}\
\epsfig{file=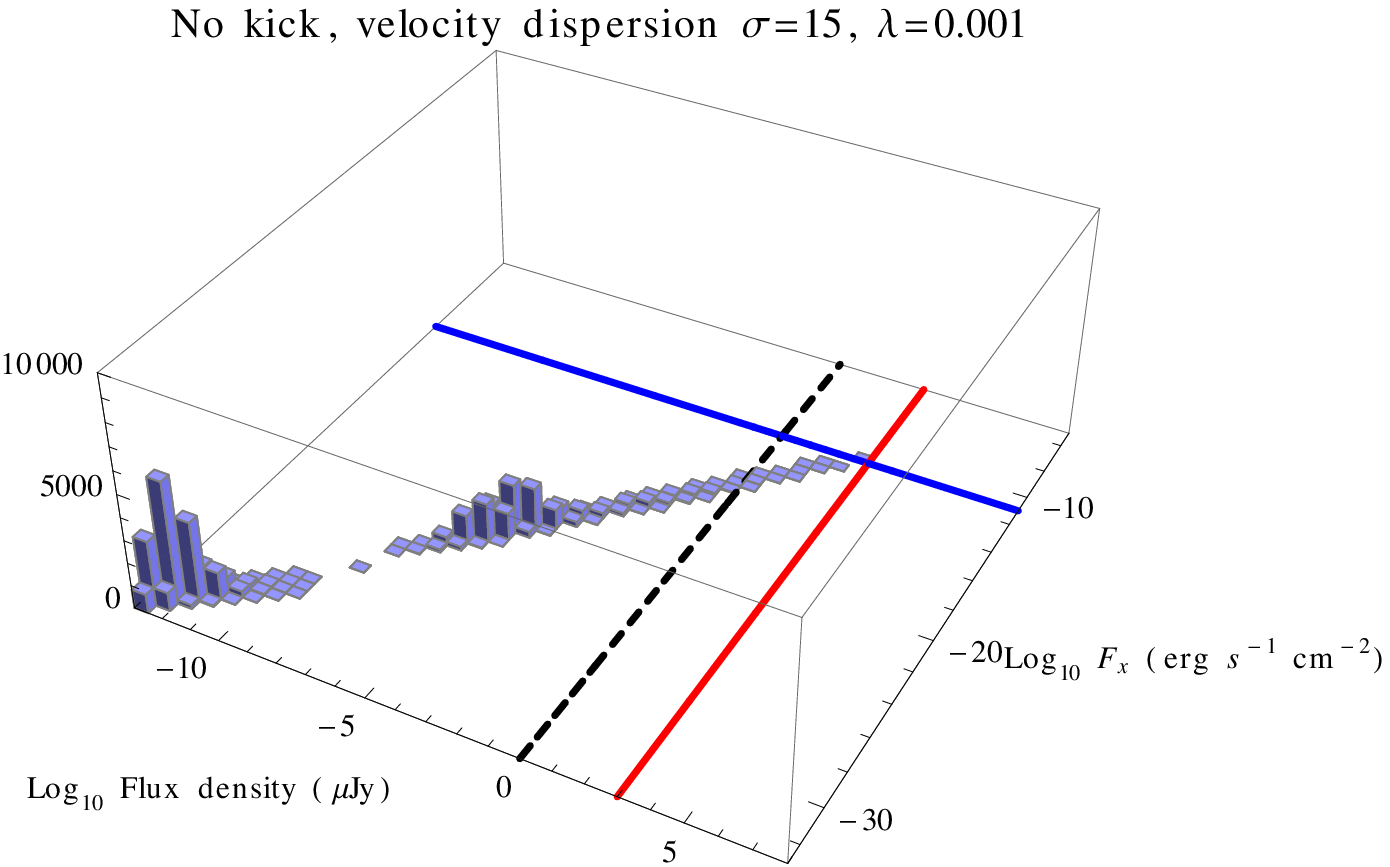,width=8cm}\quad\epsfig{file=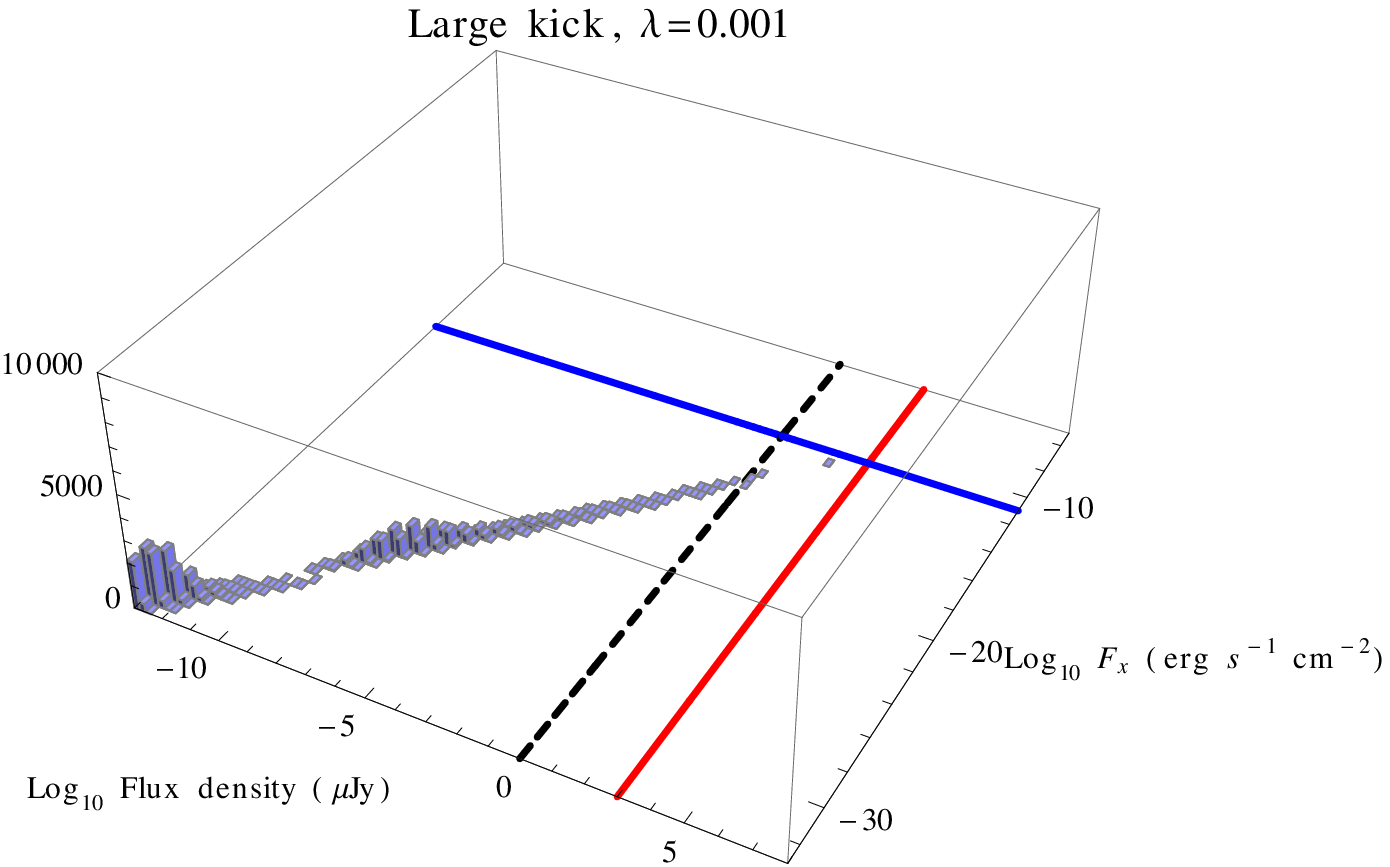,width=8cm}
\caption{
Constraints on black hole kicks and Bondi-Hoyle accretion
  parameter, $\lambda$, from our simulations. Each panel plots X-ray
  flux vs. radio flux density as a 3D histogram for a simulation
  of 34 340 IBH in a local spherical volume of radius 250pc. The strong correlation between radio and X-ray fluxes is set by the prescriptions in the text, the discontinuity at low accretion rates reflects the fact that we have set all the ISM within 70 pc to be in the hot phase.
  The
  solid lines indicate approximate source detection limits from the
  FIRST and NVSS radio surveys at $\sim 1$ mJy and hard X-ray limits from
  INTEGRAL IBIS/ISGRI and Swift BAT at $10^{-11}$ erg s$^{-1}$. The dashed lines indicate the predicted future sensitivity for SKA surveys (dashed lines) of 1 $\mu$Jy. For a low velocity dispersion and no kicks, $\lambda=0.01$ (top left) corresponds to a significant (10--100) population of objects expected in the existing surveys. Larger values for $\lambda$ are essentially ruled out unless there are very large kicks, exceeding the momenta associated with neutron star kicks. The probable lack of sources in existing X-ray catalogues implies that $\lambda \leq 0.01$ or that black holes have significant kicks. For all four scenarios a population of radio sources detectable with SKA surveys is predicted. At lower values of $\lambda$ this population of IBH will remain undetected even in the era of the SKA.
  }
\label{hrxp}
\end{figure*}

\subsection{Proper motions}

We now have a set of BH, accreting from all four main phases of the
ISM, for which we have calculated the observed flux density. For each
of these, under the assumption that their 3D velocity, $v_{\rm 3D}$,
is randomly oriented in space, we can convert to a projected 2D
transverse velocity $v_t$ by simply assuming that the angle between
the 3D velocity vector and the line of sight is uniformly distributed
in cosine. The relation simplifies to:

$v_t = \sqrt(1-x^2) v_{\rm 3D}$

\noindent
where $x$ is randomly distributed in the range $0 \leq x \leq 1$.

For a complete simulation of the proper motion we would need to add
to these proper motions a component associated with the differential rotation of the galaxy
(Oort 1972). On the scale of the local sphere simulated here, these are
minor effects, with a maximum amplitude of 7.5 km s$^{-1}$, and
resulting maximum proper motion of $\sim 6$ mas yr$^{-1}$.  However,
at distances above about 500 pc they can come to dominate if the IBH
are born with low velocities. Since our key results are going to be
dominated by much more nearby sources we do not apply these components.

Under these assumptions we have, finally, a set of objects for which we have X-ray and radio fluxes and proper motions. We are able to compare these to the capabilities of current and future X-ray and radio telescopes including.

\section{Results}
\label{results}

In this section we shall highlight the key results from our
simulations. Firstly, we tackle the issue of the two key parameters,
namely the IBH velocity distribution and the effciency of Bondi-Hoyle
accretion.

\subsection{Limits on black hole kicks and Bondi-Hoyle efficiency}

We can place limits on the Bondi-Hoyle efficiency parameter $\lambda$,
by comparing our results with all-sky hard X-ray and radio surveys (hard X-rays,
above a few keV, are necessary, because in the highest accretion rate
environments there can be a lot of local absorbing column
density). The 4th INTEGRAL IBIS/ISGRI soft gamma-ray survey catalog
(17--100 keV; Bird et al. 2010) and the 54 month Swift-BAT hard X-ray catalogue
(15--150 keV; Cusumano et al. 2010) both reach limits (for $\geq 90$\% sky
coverage) of around $10^{-11}$ erg s$^{-1}$. At these flux limits,
each survey has a small number of genuinely unidentified sources, but probably not an
entire missing population. The IBIS/ISGRI catalogue gives a relatively high
figure of 29\% unidentified objects of the $\sim 700$ in the catalogue, but
notes that this is mostly due to the poor angular resolution at hard X-rays
making cross-identification with other wavelengths difficult.
In the coming decade, these limits are only likely to be superceded by the hard component of the eROSIAT surveys (Merloni et al. 2012).

In the radio band, the most comprehensive surveys are FIRST (Becker, White \& Helfand 1995) and NVSS (Condon et al. 1998), performed with the VLA. These surveys have very
large source lists, over one million catalogued objects in NVSS to a completeness limit of $\sim 2.5$ mJy, but they do not have any attempted test for proper motion based on cross-catalogue comparisons. Without the proper motions it is almost impossible to attempt to find a small population of faint objects such as those discussed here. These  existing X-ray and radio survey limits are indicated by the solid lines in Fig \ref{hrxp}, where they are compared to a set of our simulations.

The current constraints combined with future survey limits for the SKA turn out to be very interesting when compared to our simulations. We can set limits to the boundaries of interesting parameter space by noting that for $\lambda = 10^{-4}$ and below none of these IBH are likely to ever be detectable. We focus instead on the four scenarios ($\lambda = 10^{-2},10^{-3}$) $\times$ (no kick, large kick). The largest fluxes are of course produced for ($\lambda = 10^{-2}$, no kick) which predicts 10--20 X-ray sources above the INTEGRAL-IBIS / Swift-BAT survey limits within 250 pc, corresponding to a possible total population of $\sim 100$ objects in the catalogues. This is likely to exceed the maximum number of IBH which could be `hidden' in these surveys, as discussed above. For the other three scenarios no significant population of X-ray objects would be present in the existing X-ray surveys. Considering the radio fluxes, only the ($\lambda = 10^{-2}$, no kick) scenario corresponds to any sources which could already be in the NVSS or FIRST catalogues (and then only a handful). However, with two orders of magnitude greater sensitivity, all four scenarios predict populations of radio sources which could be detectable with the SKA (dashed lines in Fig \ref{hrxp}).

\section{Detection via proper motions}

As well as radio and X-ray fluxes, the simulation delivers the proper
motions of the IBH population on the sky. In the radio band, source counts at sub-mJy flux densities are dominated by active
galactic nuclei, starburst and other types of galaxies (e.g. Seymour
et al. 2008; Padovani 2011), which will have unmeasurably small proper
motions (smaller than those estimated here for the nearby IBH by several
orders of magnitude). Therefore any faint radio source with high proper motion is likely to be of interest. Fig 4 presents the simulated population of objects as a function of both their radio flux and proper motions. In the following section we explore whether the Square Kilometer Array (Carilli \& Rawlings 2004; Hall 2004) would be able to identify these IBH via their proper motions as part of its wide field surveys.

\begin{figure*}
\epsfig{file=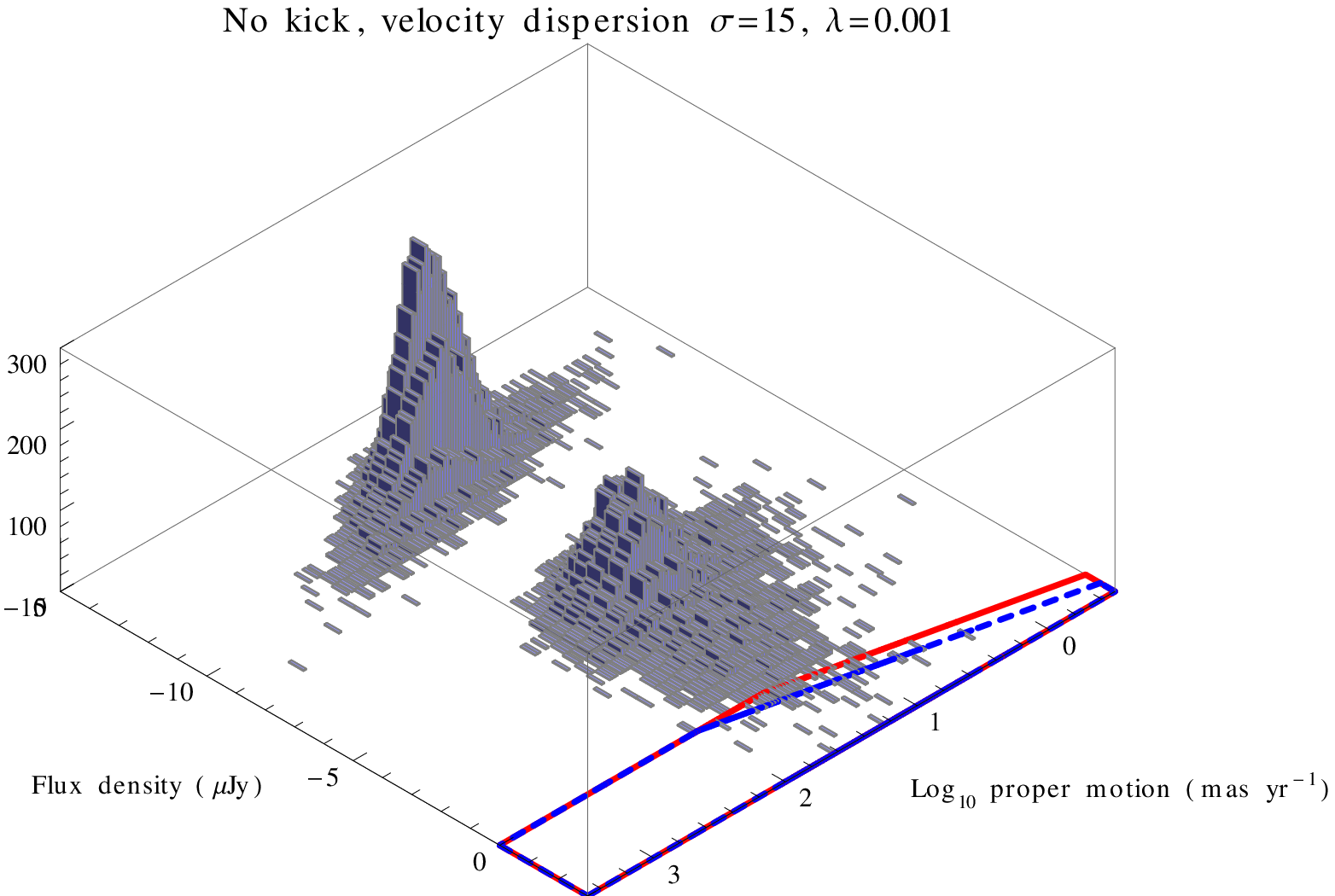,width=8cm}\quad\epsfig{file=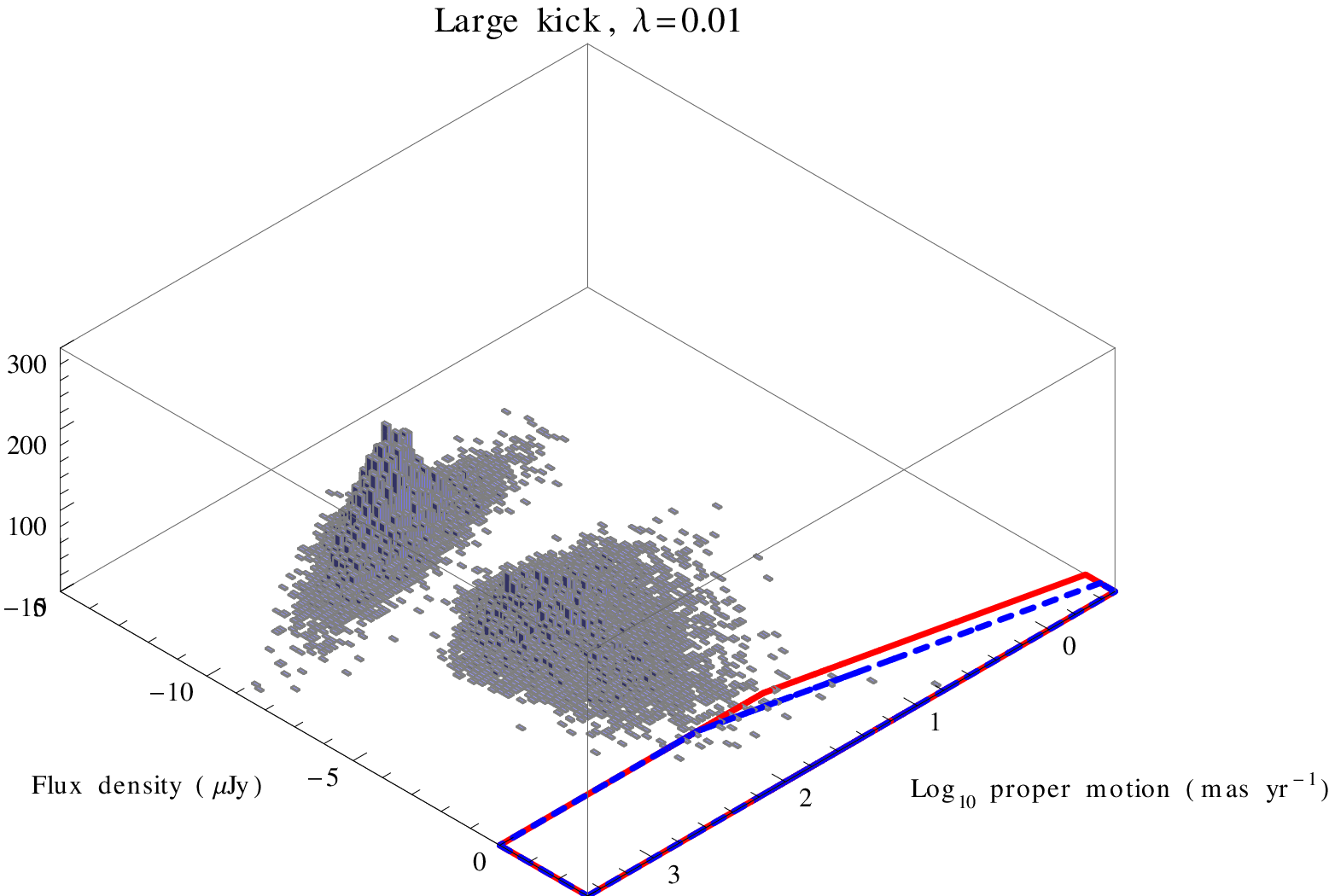,width=8cm}
\caption{Proper motion and flux for the simulated population of IBH, for the two plausible scenarios identified earlier, namely a low velocity dispersion combined with a low Bondi accretion efficiency $\lambda$, or a high kick velocity and a large $\lambda$.} 
\label{pr3d}
\end{figure*}

\subsection{Comparison with SKA simulations}

We demonstrate the feasibility of detecting IBH with the final dish component of the Phase 2 Square Kilometre Array (SKA$_{2}$) by computing a full imaging simulation based on a hypothetical multi-epoch survey. This requires three components, namely a model for the sky, a description of the observing programme and the specifications of the instrument. To perform a SKA$_{2}$ simulation in the truest sense of the word is computationally not feasible, thus several simplifications are made which are described and justified in the sections that follow.

\subsection{Assumed SKA$_{2}$ specifications}
\label{sec:ska2specs}

Our assumed specifications of the SKA$_{2}$ are drawn from those published by Dewdney et al. (2010) and are summarised in Table \ref{tab:ska2specs}. The array consists of 243 stations, each of which is assumed to consist of 12 dishes with diameters of 15~m equipped with single-pixel feeds. These dishes are beamformed at the station level and these 243 beamformed signals are then cross-correlated. This approach is analogous to that of the LOFAR telescope where signals from individual dipoles are beamformed at each station before being transmitted to the central correlator.

The stations are arranged according to four zones, each of which has a percentage of the collecting area within a certain radius: core (0~$<$~r~$<$~0.5~km, 20\%), inner (0.5~$<$~r~$<$~2.5~km, 30\%), mid (2.5~$<$~r~$<$~180~km, 30\%) and remote (180~$<$~r~$<$~3000~km, 20\%). Obviously the final placement of the dish stations within the SKA will be strongly influenced by the geography of the African host nations. A generic configuration is adopted here, featuring a random distribution in the core (with constraint to avoid station overlap), five log-spiral arms describing the inner and mid regions, and a log-spacing in radius with a random azimuthal placement for the remote stations.

\begin{table}
\centering
\caption{Assumed and derived parameters of the dish component of the Phase 2 SKA.\label{tab:ska2specs}}
\begin{tabular}{lllllll}
\hline
Number of stations ($n_{s}$)                    & 243 \\
Dishes per station                              & 12 \\
Total number of dishes                          & 2916 \\
Number of baselines                             & 29403 \\
Maximum baseline				& 2961.7 km\\
Dish diameter                                   & 15 m \\
System temperature ($T_{sys}$)                  & 30 K \\
Aperture efficiency ($\eta_{A}$)                & 0.7 \\
Correlator quantization efficiency ($\eta_{Q}$) & 0.9 \\
Instantaneous bandwidth ($\Delta \nu$)          & 1 GHz \\
\hline
\end{tabular}
\end{table}

\subsection{Sky model and survey strategy}
\label{sec:skymodel}

For the idealised sky we use the simulated catalogue of extragalactic radio continuum sources described by Wilman et al. (2008), selecting all the sources in a randomly chosen 3$\times$3 arcminute patch of the simulated sky. The simulation reaches a flux limit of 10 nJy and models sources with a combination of discrete points and extended two-dimensional Gaussians. A list of sources is extracted using the online SQL interface\footnote{\tt http://s-cubed.physics.ox.ac.uk} to the SKA Simulated Skies database and to this catalogue we add a point source near the centre of the field representing an IBH. The radio flux density of the IBH is assumed to be 10~$\mu$Jy, and it has a proper motion corresponding to a projected angular motion of 0.2 arcseconds year$^{-1}$. Including the IBH the model radio sky in these 9 arcmin$^{2}$ contains 1915 discrete radio sources.

We assume that the SKA$_{2}$ observes the patch of sky containing the IBH as part of a moderately deep sky survey at 1--2 GHz, with a total on-source time of 3 hours per pointing. Each pointing consists of six 30-minute scans within an hour angle spread of eight hours. This same survey is then repeated precisely one year later, although the starting hour angle and the scan separation are shifted in order to change the point-spread function\footnote{During an aperture synthesis observation each baseline measures a single component in the Fourier domain of the sky (the $uv$-plane) each time the correlator performs an integration. The rotation of the Earth changes the projection of each baseline on the sky and imparts an elliptical locus to its measurements in the $uv$-plane. The Fourier transform of this $uv$-plane sampling pattern (the $uv$-coverage) is the point-spread function of the observation in the image plane (also known as the dirty beam).}.

The instantaneous field of view of a 15 metre dish (see Section \ref{sec:ska2specs}) at these frequencies spans approximately 1 degree. Imaging the whole field of view at the resolution afforded by uniform weighting of the visibilities ($\sim$0.1 arcseconds) would require an image containing approximately 10 gigapixels assuming the usual three pixels per resolution element. This is clearly overkill for this simulation, hence our restriction of the field of view to 3$\times$3 arcminutes: we simply assess the IBH detection feasibility against a typical radio source background and assume that the effects of bright confusing sources outside this region can be suppressed, either by deconvolution or modelling and subtraction in the visibility domain. No attenuation of the sky due to primary beam effects is applied. 

Although any high frequency capabilities of the SKA are still uncertain, the simulation is also repeated for a 5 GHz observation using a sky model derived from the 4.8 GHz fluxes from the continuum simulation. The IBH source is assumed to be spectrally flat between ($\alpha = 0$) 1.5 and 5 GHz.

\subsection{Simulated images}

\begin{figure*}
\centering
\includegraphics[width = \textwidth]{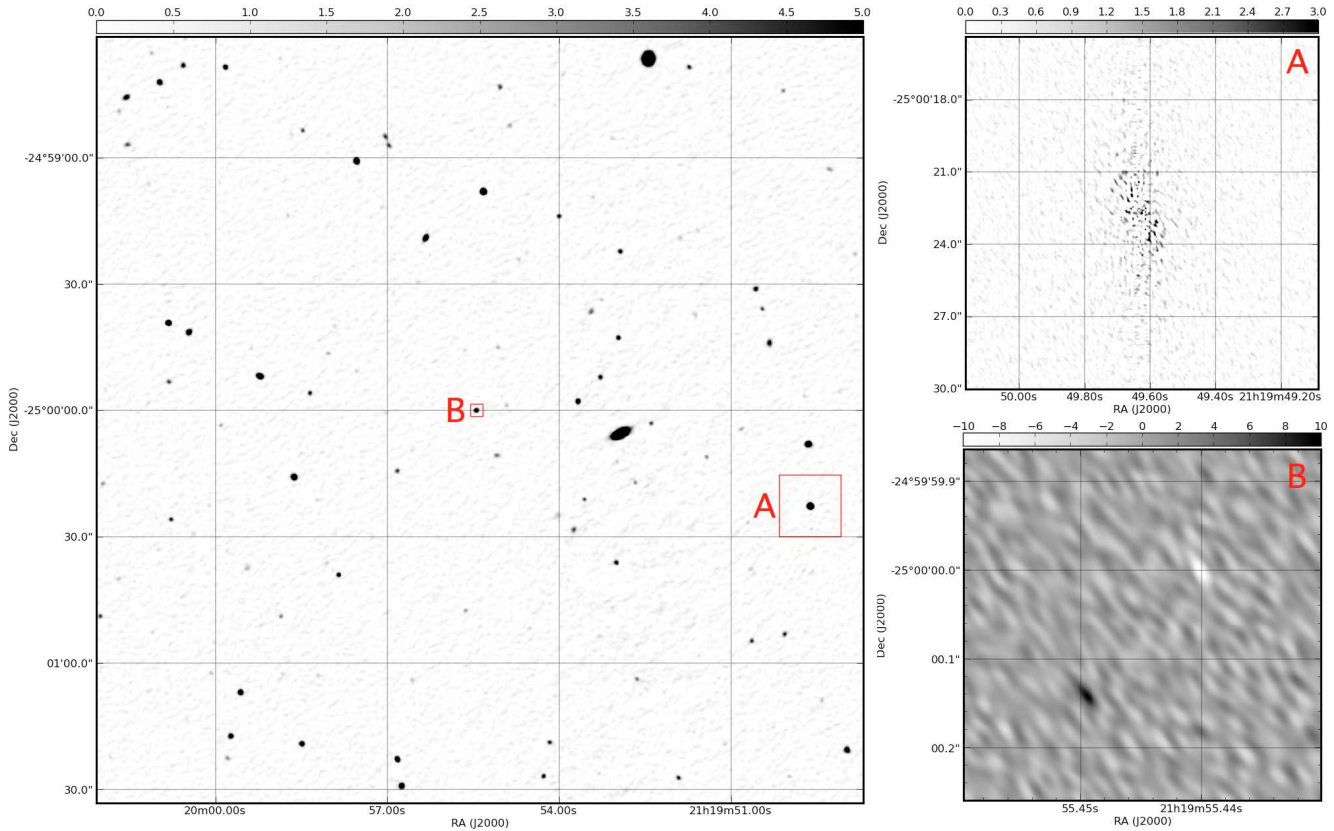}
\caption{The main panel shows one of the two simulated SKA$_{2}$ observations, at 1-2 GHz, imaged with uniform weighting, and deconvolved using the CLEAN algorithm with ten thousand clean components. Note that for clarity this image has been convolved with a 0.9 arcsecond Gaussian. Panels A and B are formed by subtracting the two simulated maps. Panel A shows the residual emission around a bright source due to incomplete removal of the differing PSFs. Panel B shows a difference image of the region around the IBH, with positive-negative emission due to the motion of the source between the observations. Pixel values are shown with a scale bar above the panel where appropriate, the units of which are in $\mu$Jy / beam. \label{fig:simulation}}
\end{figure*}

The simulated maps from the two epochs of the hypothetical survey are generated by constructing and imaging a set of model visibilities with appropriate perturbations due to thermal noise. The CASA\footnote{{\tt http://casa.nrao.edu}} package is used to generate an empty Measurement Set containing $u$, $v$, $w$, $t$, $\nu$ points consistent with the station positions and observational parameters defined in Section \ref{sec:ska2specs}. Model visibilities are then written to this database using the MeqTrees package (Noordam 2010) with the source catalogue described in Section \ref{sec:skymodel} as the sky model. For the second epoch the process is repeated with the position of the IBH is shifted by 0.2 arcseconds.

The rms noise added to each visibility and the corresponding map noise (Thompson et al. 2001) are given by
\begin{equation}
\sigma_{visibility}~=~\frac{2k_{B}T_{sys}}{\eta_{Q}\eta_{A}A\sqrt{\Delta\nu \Delta t}}
\end{equation}
and 
\begin{equation}
\label{eq:noise}
\sigma_{map}~=~\frac{w_{rms}}{w_{mean}} \frac{2k_{B}T_{sys}}{\eta_{Q}\eta_{A}A\sqrt{n_{s}\left(n_{s}~-~1\right)\Delta\nu \tau}}
\end{equation}
respectively where $A$ is the geometric collecting area of the \emph{station}, $\Delta t$ is the integration time per visibility point, $\tau$ is the total on-source time, $w_{rms}$ and $w_{mean}$ are the mean and rms values of the imaging weight factors and the rest of the symbols are as defined in Table \ref{tab:ska2specs}. For a naturally weighted map the ratio of the weight factors is 1. The ratios for the two uniformly-weighted maps we generate for this simulation are 3.04 and 3.02.

The main panel of Figure \ref{fig:simulation} shows the uniformly-weighted image of the field containing the IBH for the first simulated epoch. Note that unlike natural weighting, uniform weighting does not offer a single characteristic resolution; rather there is a trade-off between resolution and sensitivity that depends on the extent of the image. The rms noise in this image is 0.2 $\mu$Jy beam$^{-1}$, in very close agreement to the theoretical value of 0.21 $\mu$Jy beam$^{-1}$ derived from Equation \ref{eq:noise}. For clarity this image has been convolved with Gaussian beam with a full-width half-maximum spanning 0.9 arcseconds, artificially degrading the resolution of the image by a factor $\simeq$9. This image has been deconvolved non-interactively using ten thousand clean components. This blind deconvolution is the likely cause of the discrepancy between the measured and theoretical background rms levels. The IBH is near the centre, in the region marked `B'. A residual image of region B formed by subtracting the images of the 	two epochs is shown in the lower-right panel. The characteristic positive-negative emission is a result of the shifting position of the IBH.

\begin{figure}
\centering
\includegraphics[width = 0.35 \textwidth]{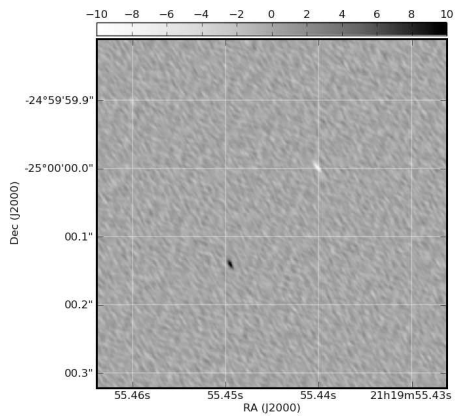}
\caption{The region of the difference image resulting from the 5~GHz sky simulation close to the IBH. Again there is corresponding negative source due to the motion of the source. The resolution advantage of the 5~GHz observations is clear, opening up the possibility of detecting sources with much lower apparent proper motions. See the text for a discussion of the associated drawbacks. Pixel values are shown in the scale bar above the figure in units of $\mu$Jy / beam. \label{fig:fivegig}}
\end{figure}

Figure \ref{fig:fivegig} shows a small area of the image resulting from the subtraction of the two simulated epochs in the 5~GHz simulation. Again there is a characteristic negative counterpart to the radio continuum source associated with the IBH. The resolution advantage afforded by moving up to 5~GHz is clear when comparing this figure to region B in Figure \ref{fig:simulation}.  The increased resolution would allow the detection of sources with much lower apparent proper motions. The disadvantage arises from the fact that the sky area covered by a single pointing of a radio telescope is proportional to $\nu^{-2}$, and a search for IBH within multi-epoch sky surveys relies on significant sky coverage. The decrease in survey speed at 5~GHz may be tempered somewhat by the fact that at higher resolution and higher frequencies it takes longer for an observation to reach the classical confusion limit, although this is not a limiting factor in the 1--2 GHz simulation presented above.

As a cautionary tale for future image plane transient searches, the panel in the upper-right of Figure \ref{fig:simulation} shows a residual image of region `A' around the brightest (50~$\mu$Jy) point source in the field. Although the sheer number and layout of the receptors of the SKA result in excellent sidelobe performance, this image is included as a reminder that caution must still be exercised. Due to the differing PSFs between observation epochs, incomplete deconvolution (as is the case here) or inadequate subtraction of bright sources in the field will leave spurious artefacts in residual images which are at similar levels to the sources we wish to detect. Intelligent algorithms for rejecting such features must be a key component when differencing radio maps to locate transient phenomena.

The fitted restoring beams in the uniformly weighted images of the two simulated survey epochs at 1--2~GHz are
0.194~$\times$~0.116 arcseconds (PA~=~56.65 degrees) and 0.189~$\times$~0.132 (PA~=~56.84 degrees). Although we have demonstrated via a full simulation the feasibility with which the SKA$_{2}$ could detect an IBH with favourable flux and motion characteristics, the results of this section can be easily adapted to analytically assess the detectability of any member of the IBH population for arbitrary observing strategies.

\section*{Discussion and Conclusions}

In this paper we have tried to see how a nearby population of black holes, which have not revealed themselves in large numbers to date in existing X-ray surveys, might be detected. The rewards of identifying the right strategy may be high, as the detection of a black hole at a distance of only a few pc would be of great interest to astrophysicists and relativists (although it is worth noting that even then, Sgr A* would have a larger projected size on the sky). 

The comparison of predicted X-ray luminosities with existing surveys reinforces the conclusions of Perna et al. (2003), who considered the population of nearby neutron stars, that the accretion rates onto nearby compact objects must be orders of magnitude lower than the na\"ive Bondi-Hoyle predictions. The prospects for future deeper hard X-ray surveys which might find this population of faint objects are not particularly good, and it may be that the limits from INTEGRAL-IBIS and Swift-BAT remain the state of the art for some time. The best prospect in the near future is the hard band (2--10 keV) component of the eROSITA survey (Merloni et al. 2012) which, with a possible order or magnitude improvement in sensitivity, could detect the bright end of the population.
However, as we have shown, if isolated black holes have a similar relation between accretion rate and radio luminosity as other low Eddington ratio black holes, a significant population of them may be detectable with future radio telescopes, most notably the SKA. Of course this is the crucial assumption -- whether or not sources undergoing Bondi accretion from the ISM by a black hole with relatively high velocity will have accretion flow axisymmetic enough to produce jets as powerful as those produced by X-ray binaries and low luminosity AGN (the latter incorporating a mass scaling, e.g. Gultekin et al. 2009). We note that very recently Barkov, Khangulyan \& Popov (2012) have argued that IBH can produce jets, even without the formation of a large accretion disc, via the Blandford-Znajek (1977) mechanism (although of course whether or not the B-Z mechanism actually operates in nature is unclear -- see Fender, Gallo \& Russell (2010) and Narayan \& McClintock (2012)). It has further been argued that the Bondi prescription itself is not appropriate in the presence of angular momentum (Power, Nayakshin \& King 2011), and the accretion flows produced in numerical simulations can exhibit complex behaviour (e.g. Blondin \& Raymer 2012).

Nevertheless, if the IBH really are radio sources with the predicted luminosities, then it should be possible to pick them out of future repeated wide-field, deep radio surveys, as faint radio sources with measurable proper motions, setting them apart from the rest of the sub-mJy population which is dominated by distant AGN and starburst galaxies with no measurable proper motions. The SKA as currently envisaged, is clearly capable of making such measurements, but whether or not it would is currently not clear. In the simulation presented here, a one degree field (at 1.5 GHz) is observed with single pixel feeds (SPFs) for three hours, and easily detects the high proper motion IBH. In our simulations, between 10--100 IBH are detectable via their proper motions, which implies that at least a quarter of the sky would need to be imaged twice in order to detect a handful of IBH. This implies the observation of around 20 000 fields which, at 3 hr per observation, corresponds to about seven years of observing. The observation of wider fields with phased array feeds (PAFs) such as those being developed for WSRT/APERTIF and ASKAP, with a 30 deg$^{2}$ field of view, would of course dramatically reduce this time -- at the same sensitivity (i.e. bandwidth), seven years would become a few months. In reality, the bandwidth achievable by the FPAs is probably not as great as that for the SPFs, and so the survey time is likely to lie between these two extrema.

Considering again Figure 4, we can see that we are in fact more likely to detect a population of nearby IBH if they do {\em not} have very large kicks. This is because the angular resolution of the SKA (and indeed many current radio interferometers and optical telescopes) is already good enough to measure the proper motions of nearby objects with relatively small velocity dispersions, and the reduction in accretion rate due to high space velocities is much more of an issue. 
It is also worth considering that basing our assumptions about the mass distribution of IBH on that measured for X-ray binaries may be incorrect, and that field IBH may have a large mean mass and broader mass distribution. Since the relation between radio luminosity and black hole mass is close to cubic, this could be an extremely important term.

What about optical emission? In most low-accretion rate black holes, the optical emission is
dominated by reprocessing by the outer accretion disc of X-rays from the innermost parts of the flow or by the companion star (e.g. Russell et al. 2006 and referencetherein). Given the highly uncertain geometry of Bondi accretion it seems rather unlikely that a large enough disc forms for this reprocessing to occur, and there will of course be no companion star. However, there is good evidence that flat-spectrum ($\alpha= 0$) emission from the jet (if it exists) might extend to the near-infrared or even optical bands (e.g. Corbel \& Fender 2002; Gandhi et al. 2011). Such emission could potentially be detected from a handful of nearby IBH by deep wide-field surveys such as UKIDSS (Lawrence et al. 2007) or WISE (Wright et al. 2010), although finding such objects in these surveys without any additional information seems highly unlikely.

Finally, is there any chance to find these IBH now in any existing data? It is possible that a small number of these objects are there in existing X-ray or radio studies but, as noted earlier, the poor angular resolution in the hard X-ray band makes it very hard to identify the radio counterparts of faint X-ray sources (and vice versa). However, the numbers of IBH also reveal the possibility that there may be a black hole (or, even more likely, neutron star) accreting from the very nearest clouds in the ISM (those within 15 pc). Careful X-ray and radio observations of these clouds just might reveal the presence of such an object.

\section*{Acknowledgements}

RPF would like to thank Ed van den Heuvel and Evan Keane for useful discussions about IBH numbers, Jon Slavin, Pris Frisch and Seth Redfield for very helpful discussions about the local ISM, and Tony Bird for discussions about INTEGRAL limits and populations.
RPF is supported in part by European Research Council Advanced Grant 267697 "4 pi sky: Extreme Astrophysics with Revolutionary Radio Telescopes.  IH
wishes to thank Rosie Bolton and Rob Millenaar for their assistance
with the generic station layouts. IH thanks SEPnet for financial
support.

\end{document}